%% file: main.tex
\documentclass[sigconf,nonacm]{acmart}
\input{include/pkgs}
\AtBeginDocument{%
  }
\setcopyright{none}
\author{Haocheng Xu}
\authornote{Both authors contributed equally to this work.} 
\email{haochx5@uci.edu}
\affiliation{\institution{University of California, Irvine}\city{Irvine}\state{California}\country{USA}}

\author{Yu-an Chou}
\authornotemark[1]
\email{chouy10@uci.edu}
\affiliation{\institution{University of California, Irvine}\city{Irvine}\state{California}\country{USA}}

\author{Bhardwaj Bhat}
\email{bhardwb@uci.edu}
\affiliation{\institution{University of California, Irvine}\city{Irvine}\state{California}\country{USA}}

\author{Zhiheng Chen}
\email{zhihenc5@uci.edu}
\affiliation{\institution{University of California, Irvine}\city{Irvine}\state{California}\country{USA}}

\author{Ye Qiao}
\email{yeq6@uci.edu}
\affiliation{\institution{University of California, Irvine}\city{Irvine}\state{California}\country{USA}}

\author{Yifan Zhang}
\email{yifanz58@uci.edu}
\affiliation{\institution{University of California, Irvine}\city{Irvine}\state{California}\country{USA}}

\author{Leyao Han}
\email{leyaoh5@uci.edu}
\affiliation{\institution{University of California, Irvine}\city{Irvine}\state{California}\country{USA}}

\author{Saptarshi Mitra}
\email{saptarm@uci.edu}
\affiliation{\institution{University of California, Irvine}\city{Irvine}\state{California}\country{USA}}

\author{Sitao Huang}
\email{sitaoh@uci.edu}
\affiliation{\institution{University of California, Irvine}\city{Irvine}\state{California}\country{USA}}
\begin{document}


\title{LowRank-SSM: Hardware-Software Co-Design for Rank-Reduced Mamba Acceleration on FPGA}

\begin{abstract}
State Space Models (SSMs) such as Mamba and Mamba-2 achieve linear-time autoregressive inference, making them attractive for latency-sensitive and resource-constrained deployment. Yet every generated token still reads the full input and output projection weights of every layer from off-chip memory, and on an FPGA these two projections take the majority of a layer's cycles. Existing accelerators lower this cost by quantizing the weights or by speeding up the scan, but none reduce the number of weight elements a token must read, and a low-rank factorization on its own does not help: without a datapath built for it, the two-stage product takes as long as the dense one.

We present \textbf{LowRank-SSM}, a hardware-software co-design framework that jointly addresses three constraints of rank-reduced Mamba inference: where rank should be spent, how the resulting factors should execute, and how reduced weight traffic should be delivered by the FPGA memory system. A sensitivity-guided per-projection search allocates rank under a perplexity ceiling while restricting compression to Mamba-2 operands that tolerate approximation. A rank-configurable projection datapath maps both factor stages onto the existing GEMV engines, keeping the intermediate vector on chip and changing rank through loop bounds and weight images rather than bitstream recompilation. A platform-aware NoC/AXI memory subsystem then balances and prefetches weight streams to sustain 42\,GB/s. Together, these software and hardware choices convert structural rank reduction into measured latency reduction: in our ablation, 93\% of the projection traffic removed by rank reduction appears as saved token latency. With a rank schedule found under a 5\% perplexity ceiling, which costs 1.6 points of average zero-shot accuracy, the complete accelerator decodes Mamba2-2.7B at 16.79 tokens/s and 0.41\,J per token.

\end{abstract}
\maketitle
\input{sections/01-Introduction}

\input{sections/02-Background}
\input{sections/03-SoftwareDesign}
\input{sections/04-HardwareDesign}

\input{sections/05-Experiments}
\input{sections/06-Conclusion}

\bibliographystyle{unsrt}
\bibliography{
  ref/model,
  ref/task,
  ref/quant
}

\end{document}

%% file: include/pkgs.tex
\usepackage{amsmath,amssymb}
\usepackage{graphicx}
\usepackage[dvipsnames]{xcolor}
\usepackage{fancyhdr}

\usepackage{array}
\usepackage{hyperref}
\usepackage{arydshln}
\usepackage{multirow}
\usepackage{booktabs}
\usepackage{xspace}
\usepackage[normalem]{ulem}
\usepackage{algorithm}
\usepackage{pifont}
\usepackage{threeparttable}
\usepackage{siunitx}
\usepackage[table]{xcolor}
\usepackage{subcaption}
\usepackage{algpseudocode}
\usepackage{algorithm}


%% file: sections/01-Introduction.tex

\section{Introduction}
\label{sec:intro}

\begin{figure}[t]
\centering
\includegraphics[width=\columnwidth]{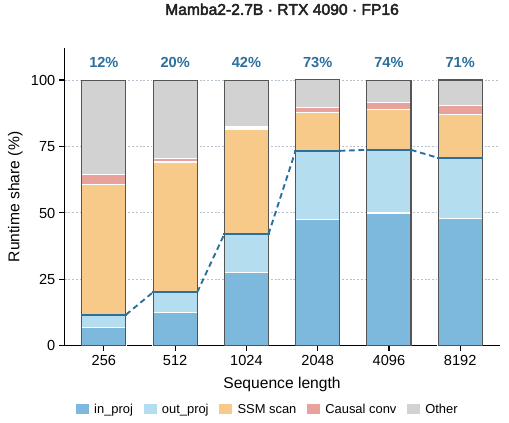}
\caption{Percentage Breakdown of Mamba2-2.7B~\cite{Dao2024}}
\label{fig:breakdown}
\vspace{-3mm}
\end{figure}

State Space Models (SSMs), particularly the Mamba family of selective SSMs~\cite{Gu2021, Gu2023, Dao2024}, have emerged as a practical alternative to Transformer-based LLMs for latency-sensitive and efficient deployment. By maintaining a fixed-size recurrent state rather than a growing key-value cache, Mamba achieves 4--5$\times$ higher autoregressive throughput than comparably sized Transformers~\cite{Gu2023}, making FPGA-based deployment attractive. SSM deployment on edge and its superiority in supporting larger input context ($\sim$4$\times$ of similar-sized Transformers) has been explored in~\cite{mitra2025characterizing}. The hardware-aware algorithmic design of Mamba rests on the critical empirical assumption that memory I/O during the SSM recurrence itself is the bottleneck. Figure~\ref{fig:breakdown} shows the runtime breakdown of Mamba2-2.7B~\cite{Dao2024}, profiled on NVIDIA 4090. The share of the input and output projections grows with sequence length as the kernels become bandwidth-bound, and from 2048 tokens on they account for 71--74\% of the time, while the SSM scan, the operation that defines the architecture, takes less than 15\%.

The true bottleneck is memory bandwidth against the projection weights, not arithmetic in the recurrence. Prior SSM accelerators~\cite{marca24,marcav2,wei2025,wang2025fastmambahighspeedefficientmamba,zhong2025specmambaacceleratingmambainference} address this either by optimizing the scan's element-wise operations or by reducing the arithmetic cost of streaming the projection weights through quantization. Both strategies leave the number of weight elements unchanged: quantization shrinks each weight, but a quantized matrix is still a full-rank matrix, and every element is still read from off-chip DRAM for every generated token. No prior accelerator reduces the size of the projection matrices themselves.

Rank reduction provides a structural way to reduce the number of weight elements transferred for every generated token: substituting a projection weight $W \in \mathbb{R}^{m\times n}$ with two smaller factors of rank $r$. This approach decreases on-chip storage and DDR bandwidth from $mn$ to $r(m+n)$ bytes per token, achieving savings that quantization cannot equal at the same precision.

However, our key observation is that this saving is useful only when three decisions are made together. First, rank must be allocated across projections without violating model-quality constraints. Second, the accelerator must execute the resulting two-factor cascade without reintroducing its cost through extra engines, off-chip intermediate storage, or pipeline stalls. Third, the memory system must deliver the irregular per-layer weight streams at sufficient and balanced bandwidth; otherwise, reduced matrix size does not translate into reduced token latency.

In this paper, we present LowRank-SSM, a co-design framework that treats projection rank as an explicit hardware design variable.
We make three contributions:
\begin{enumerate}
\item \textbf{Quality-constrained rank allocation.} We identify the Mamba-2 operands that tolerate approximation and perform per-projection greedy rank allocation under a user-specified perplexity ceiling. The resulting schedule is compatible with the accelerator's rank granularity and determines where structural compression is actually worthwhile.
\item \textbf{A co-designed memory subsystem for rank-reduced projections.} We use striped weight streams, balanced AXI master placement, embedded scales, and output-projection prefetching to sustain the bandwidth required by heterogeneous rank schedules, reaching 42\,GB/s.
\item \textbf{A rank-configurable projection datapath.} Both factor stages reuse the existing GEMV engines and on-chip buffers, with rank entering as a loop bound rather than a hardware specialization. Consequently, a new rank schedule changes the weight image and rank table without requiring a new bitstream.
\end{enumerate}

The resulting co-designed accelerator decodes Mamba2-2.7B~\cite{Dao2024} at 16.79 tokens/s and 0.41\,J per token on a Versal VC1902 at 230\,MHz, 4.65$\times$ the throughput and 1.68$\times$ the energy efficiency of LightMamba~\cite{wei2025} and 3.16$\times$ the throughput and 1.38$\times$ the energy efficiency of FastMamba~\cite{wang2025fastmambahighspeedefficientmamba}.

%% file: sections/02-Background.tex
\section{Background and Motivation}


\subsection{Selective SSMs and Projection Bottleneck}
\label{sec:background}

\begin{figure}
    \centering
    \includegraphics[width=1.0\linewidth]{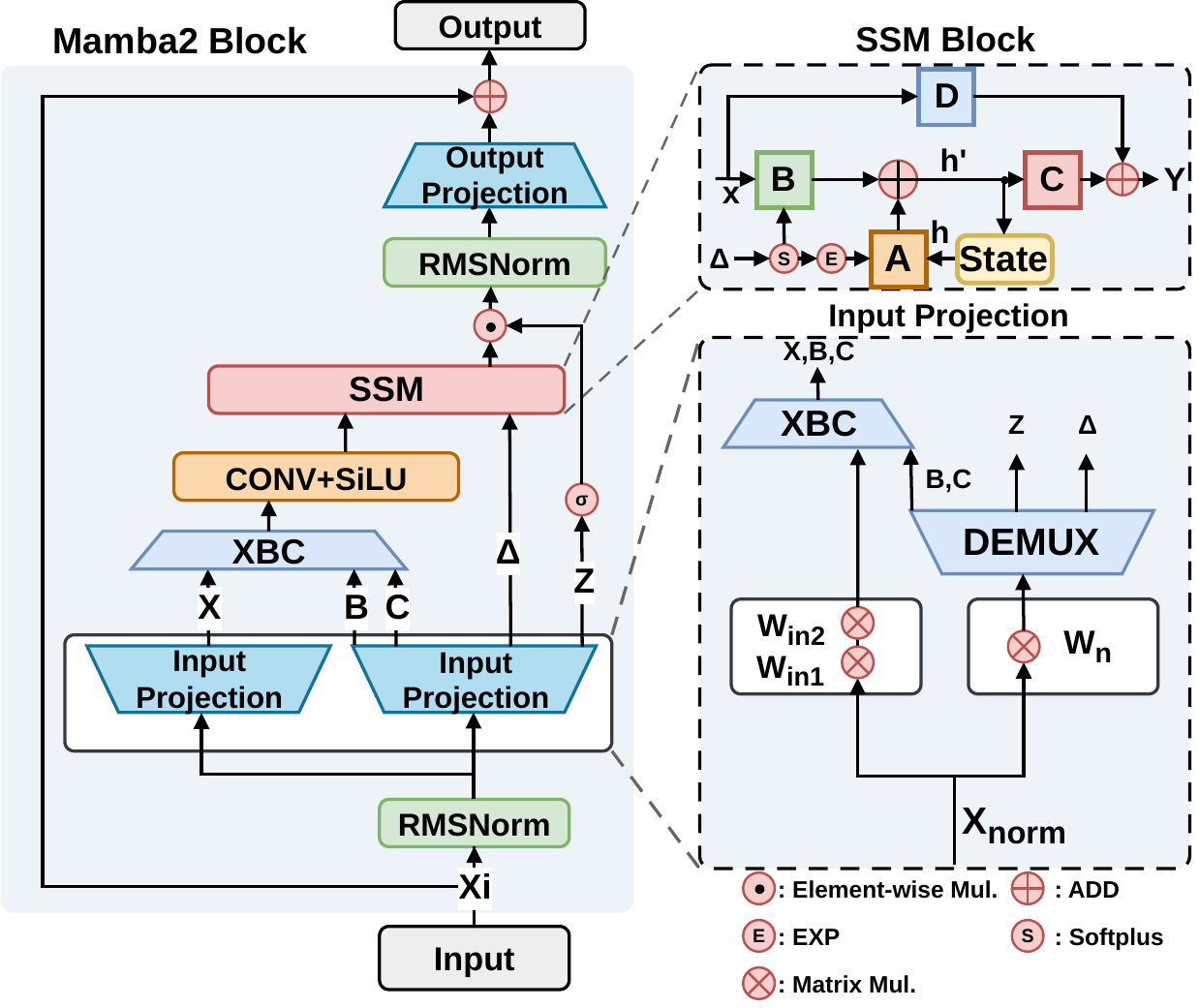}
    \caption{Mamba2 Block with LowRank}
    \label{fig:Mamba2}
\end{figure}

An SSM maps an input sequence $x_t \in \mathbb{R}^D$ to outputs through a fixed-size latent state $h_t \in \mathbb{R}^{N \times d_{\text{inner}}}$ via the recurrence
\begin{equation}
  h_t = \bar{A} h_{t-1} + \bar{B} x_t, \qquad
  y_t = C h_t + D x_t,
  \label{eq:ssm}
\end{equation}
where $\bar{A} = e^{\Delta A}$ and $\bar{B} = (\Delta A)^{-1} (e^{\Delta A} - I)\Delta B$ are input-dependent discretizations of the continuous-time matrices.  Mamba~\cite{Gu2023} makes $\Delta$, $B$, and $C$ functions of the current token, enabling content-dependent state updates; Mamba-2~\cite{Dao2024} further improves hardware efficiency via the structured state-space duality (SSD) formulation. As shown in Figure ~\ref{fig:Mamba2}, each Mamba block partitions its computation into four stages: (i)~an \emph{input projection} that maps the normalized token $x \in \mathbb{R}^D$ to five operands $[X, Z, \Delta, B, C]$; (ii)~a causal depthwise \texttt{conv1d}; (iii)~the selective scan of Eq.~\eqref{eq:ssm}; and (iv)~an \emph{output projection} back to residual space. 
In Mamba2-2.7B, the parameters are set as follows: $D = 2560$, $d_{\text{inner}} = 5120$, and $N = 128$.


Despite the $\mathcal{O}(1)$-per-step recurrence, autoregressive decode is memory-bandwidth bound. Each token requires loading the entire projection weight matrices from off-chip DRAM: for Mamba2-2.7B the input projection of one layer is a $2560\times10{,}576$ matrix and the output projection $5120\times2560$, 40.2 million weights per layer and 2.57\,GB per token in INT8 across the 64 layers. Figure~\ref{fig:breakdown} shows that at sequence length 2048, \texttt{in\_proj} and \texttt{out\_proj} together consume 73\% of the per-token time while the SSM scan takes 11\%. The bandwidth requirement scales as $\mathcal{O}(D^2)$ per layer and exceeds the DDR4 bandwidth of commodity FPGA boards at sustained generation.

\subsection{Low-Rank Decomposition}
\label{sec:bg_lowrank}

Truncated SVD provides a principled route to reducing the weight dimension.  Given $W \in \mathbb{R}^{m \times n}$, the rank-$r$ approximation $\hat{W}_r = B_r A_r$ (where $A_r \in \mathbb{R}^{r\times n}$, $B_r \in \mathbb{R}^{m \times r}$) replaces a single MVM with a two-stage cascade:
\begin{equation}
  z = A_r x \in \mathbb{R}^r, \qquad y = B_r z \in \mathbb{R}^m,
  \label{eq:factored}
\end{equation}
reducing weight storage from $mn$ to $r(m{+}n)$ elements.
The crossover rank below which factorization yields net savings is $r^* = mn/(m{+}n)$.  For \texttt{in\_proj}'s $W_X$ sub-matrix ($m{=}5120$, $n{=}2560$), any $r < 1707$ saves storage; whether the accuracy cost is acceptable is what Section~\ref{sec:rank_DSE} decides. The smaller sub-matrices also exhibit lower dynamic range than the full product matrix, making them more amenable to INT8 or INT4 quantization, which is a benefit observed in Transformer SVD compression~\cite{yuan2025asvdactivationawaresingularvalue,wang2025svdllmtruncationawaresingularvalue}.

\subsection{Related Work}
\label{sec:related_work}

\textbf{SSM accelerators.}
(a) \emph{Projection quantization}: LightMamba~\cite{wei2025} proposes rotation-assisted post-training quantization to suppress activation outliers, reducing the majority of projection computation to INT4; FastMamba~\cite{wang2025fastmambahighspeedefficientmamba} combines Hadamard quantization with a dedicated nonlinear approximation unit for the SSM block.
(b) \emph{SSM recurrence optimization}: MARCA~\cite{marca24} and its successor MARCA-v2~\cite{marcav2} identify element-wise SSM as the dominant bottleneck at long sequence lengths and
propose reconfigurable datapath units and column-wise activation sparsity to address it.
(c) \emph{Decoding throughput}: SpecMamba~\cite{zhong2025specmambaacceleratingmambainference}
introduces speculative decoding for Mamba on FPGA, addressing the hidden-state backtracking challenge unique to recurrent architectures.
All of these approaches operate on dense projection matrices of full rank. Quantization reduces the bytes per weight, but none of these designs reduces the number of weights that must be read from off-chip memory per token, since the matrices are never structurally compressed.


\textbf{LLM Weight SVD.}
ASVD~\cite{yuan2025asvdactivationawaresingularvalue} and SVD-LLM~\cite{wang2025svdllmtruncationawaresingularvalue} demonstrate effective post-training rank reduction for Transformer attention and MLP weights, achieving 10\% to 30\% parameter reduction without fine-tuning by scaling or whitening weights before SVD truncation.
However, neither addresses the SSM-specific structure where only a \emph{subset} of the input-projection outputs ($X$) benefits from rank reduction while others ($Z$, $B$, $C$, $\Delta$) require full-rank fidelity.  More critically, neither pairs the decomposition with a custom hardware datapath. On a generic accelerator the two-stage MVM of Eq.~\eqref{eq:factored} replaces one GEMM with two sequential GEMMs, and the bandwidth savings disappear into pipeline overhead unless the intermediate vector $z$ is held on-chip and both stages are fused into a single dataflow
region.

\textsc{LowRank-SSM} addresses both limitations jointly: it
identifies which projection operands tolerate rank reduction, allocates rank across layers under a perplexity ceiling at the accelerator's rank granularity, and co-designs a factored projection datapath for bandwidth savings on the FPGA.

%% file: sections/03-SoftwareDesign.tex
\section{Rank-Reduction Design Space Exploration}
\label{sec:rank_DSE}
Realizing the bandwidth savings of low-rank projection requires answering three questions before a single weight is factored: which operands in each layer should be compressed, how far each may be truncated under an accuracy budget, and which truncation rule to use. This section addresses each in turn, and then shows how the result composes with quantization.

\subsection{Operand Selection}
\label{sec:operands}
The input projection maps a normalized token $x\in\mathbb{R}^{D}$ to five distinct operands: $X,Z\in\mathbb{R}^{d_{\mathrm{inner}}}$, $\Delta\in\mathbb{R}^{d_\Delta}$ and $B,C\in\mathbb{R}^{N}$. These operands play structurally different roles in the Mamba block and therefore have very different sensitivities to rank reduction. $Z$ is injected directly at the output gate of the SSM, where it multiplies the scan output element-wise before the residual add; any approximation error in $Z$ propagates to the residual stream unattenuated. $B$, $C$ and $\Delta$ are the state parameters: errors here corrupt the recurrent dynamics and accumulate over the sequence. Their output dimensions are also small ($d_\Delta=80$, $N=128$), so they contribute little to the projection cost. $X$, s large as Z at 5120 of the 10,576 rows, feeds the depthwise convolution and then the scan, where the recurrence partially absorbs approximation errors. We therefore factorize only the $W_X\in \mathbb{R}^{d_{\mathrm{inner}}\times D}$ slice and the output projection $\mathbb{R}^{D\times d_{\mathrm{inner}}}$, which produces the residual-path update. 

\subsection{Per-Projection Greedy Rank Search}
\label{sec:method:greedy}
Algorithm~\ref{alg:greedy} summarizes the procedure. Each target (the projection) starts dense. At each step, every eligible target is tried at the next lower rank of a rank ladder $\mathcal{R}$, and the candidate with the lowest WikiText-2 perplexity is accepted, subject to a ceiling $\Delta$ above the dense model’s WikiText-2 perplexity. Trying all T=128 targets at every step would cost 128 perplexity evaluations per accepted step, which is time-consuming and inefficient; a top-eight candidate filter, which prioritizes untried candidates and the candidates that were promising in the previous step, limits the evaluations to eight, which balances out the search time and model quality. In our design space exploration process, top-8 plus greedy produce negligible difference compared to all target search. The search is per projection: each input and each output projection receives its own rank, and nothing forces consecutive layers to share one. 
The search does not depend on how a target is factored once its rank is chosen. We use three decomposition rules as drop-in primitives: plain truncated SVD; activation-aware SVD (ASVD~\cite{yuan2025asvdactivationawaresingularvalue}, input scaling with $\alpha=0.5$); and whitened SVD (SVD-LLM~\cite{wang2025svdllmtruncationawaresingularvalue}, calibrated on WikiText-2 training activations). None requires fine-tuning, and the same schedule can be re-factored with any of them.

\begin{algorithm}[t]
\caption{Per-projection greedy rank search}\label{alg:rank-dse}
\begin{algorithmic}[1]
\Require Targets $T$, ratios $R$, PPL limit $\Delta$
\State $S \gets \mathrm{dense}$ for every target
\State $P \gets \mathrm{PPL}(S)$
\While{an eligible rank reduction exists}
  \State $C \gets \mathrm{Top8}(\mathrm{Next}(S,T,R))$
  \If{$C=\emptyset$}
    \State \textbf{break}
  \EndIf
  \For{$c\in C$}
    \State $p_c \gets \mathrm{PPL}(S\oplus c)$
  \EndFor
  \State $c^* \gets \arg\min_{c\in C} p_c$
  \If{$\Delta$ is set and $p_{c^*}>P(1+\Delta$)}
    \State \textbf{break}
  \EndIf
  \State $S \gets S\oplus c^*$
\EndWhile
\State \textbf{return} $S$
\end{algorithmic}
\label{alg:greedy}
\end{algorithm}

\subsection{Where the Rank Is Spent Matters Most}
\label{sec:searchresults}
\begin{figure}[t]
\centering
\includegraphics[width=\columnwidth]{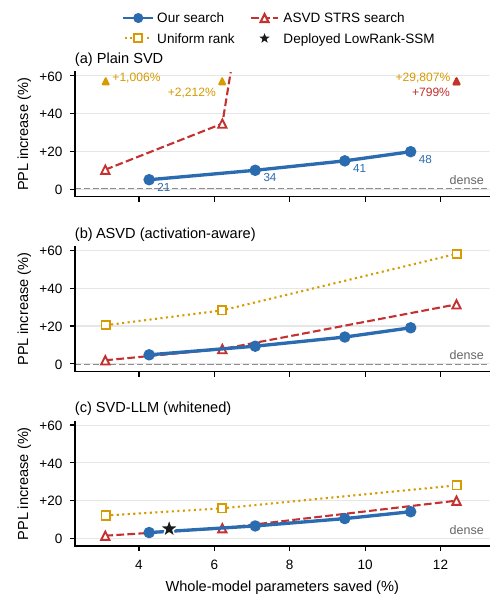}
\caption{WikiText-2 perplexity increase over the dense model versus whole-model parameters saved, for three rank allocations (curves) under three truncation rules (panels). Arrows mark points above the axis, labelled with their value; the dashed line is the dense model; the numbers on our curve in (a) are the targets compressed out of 128}
\label{fig:search}
\vspace{-3mm}
\end{figure}

Figure~\ref{fig:search} summarizes what the search buys. Each panel fixes one truncation rule and compares three ways of deciding where the rank goes: a \textbf{uniform rank} for all 128 targets, which is how SVD-LLM allocates; the sensitivity-based search that ASVD ships with, \textbf{ASVD STRS}; and \textbf{our per-projection greedy search}. Reading down a panel compares allocations under the same rule; reading across the panels shows how the same allocation fares when the factors are computed differently.
 
From design space exploration of applying different ranks to different layers of projections in Mamba2-2.7B, we found that some layers are load-bearing, and the quality of a low-rank compression is decided by whether the search algorithm finds them. As shown in Figure~\ref{fig:search}, truncating every target by the same amount (uniform rank) destroys the model with plain SVD, with a perplexity above 100 at only 3\% savings, and the more you saved, the worse it gets. Our search never touches those sensitive layers, because our search accepts a rank reduction only after measuring its effect in dropping perplexity. That alone makes plain SVD usable from 100+ ppl down to around 14 ppl (panel(a)). Similar observations can be made in panel (b) and (c): our search is better than uniform rank and ASVD-STRS, and can improve the model quality when using the same decomposition methods. 
 
Two conclusions follow. First, once the allocation is right (which rank to what layer), the truncation rule becomes a second-order choice with a predictable effect: on our search, ASVD is indistinguishable from plain SVD. Second, a schedule that protects the sensitive projections transfers across rules, while a schedule that relies on the rule does not: the ranks returned by ASVD-STRS's search work only with ASVD's factors and collapse when applied with plain SVD, whereas ours were can improve with both plain SVD and SVD-LLM. For the hardware, the practical consequence is that our schedules degrade gracefully as the saving grows, where the other two allocations bend upward, so the ceiling can be set by the accuracy a deployment needs. We deploy the $\Delta=5\%$ schedule, searched by our algorithm with SVD-LLM as decomposition, marked by the star in Figure~\ref{fig:search}.

\subsection{Quantization}
\label{sec:method:quantization}

We experimented with symmetric groupwise INT4/INT8 quantization with 128 weights per scale for low-rank factors and uncompressed projections. With INT8-only quantization, the perplexity measured almost the same against FP16 baseline. So our final hardware design pick INT8 version for deployment. We also tested alternative fixed-point formats for scales, activations, and accumulators, choosing wider intermediate accumulators to avoid premature saturation while balancing numerical accuracy against hardware cost.

%% file: sections/04-HardwareDesign.tex

\section{LowRank-SSM Hardware Design Methodology}
\begin{figure}[t]
    \centering
    \includegraphics[width=\linewidth]{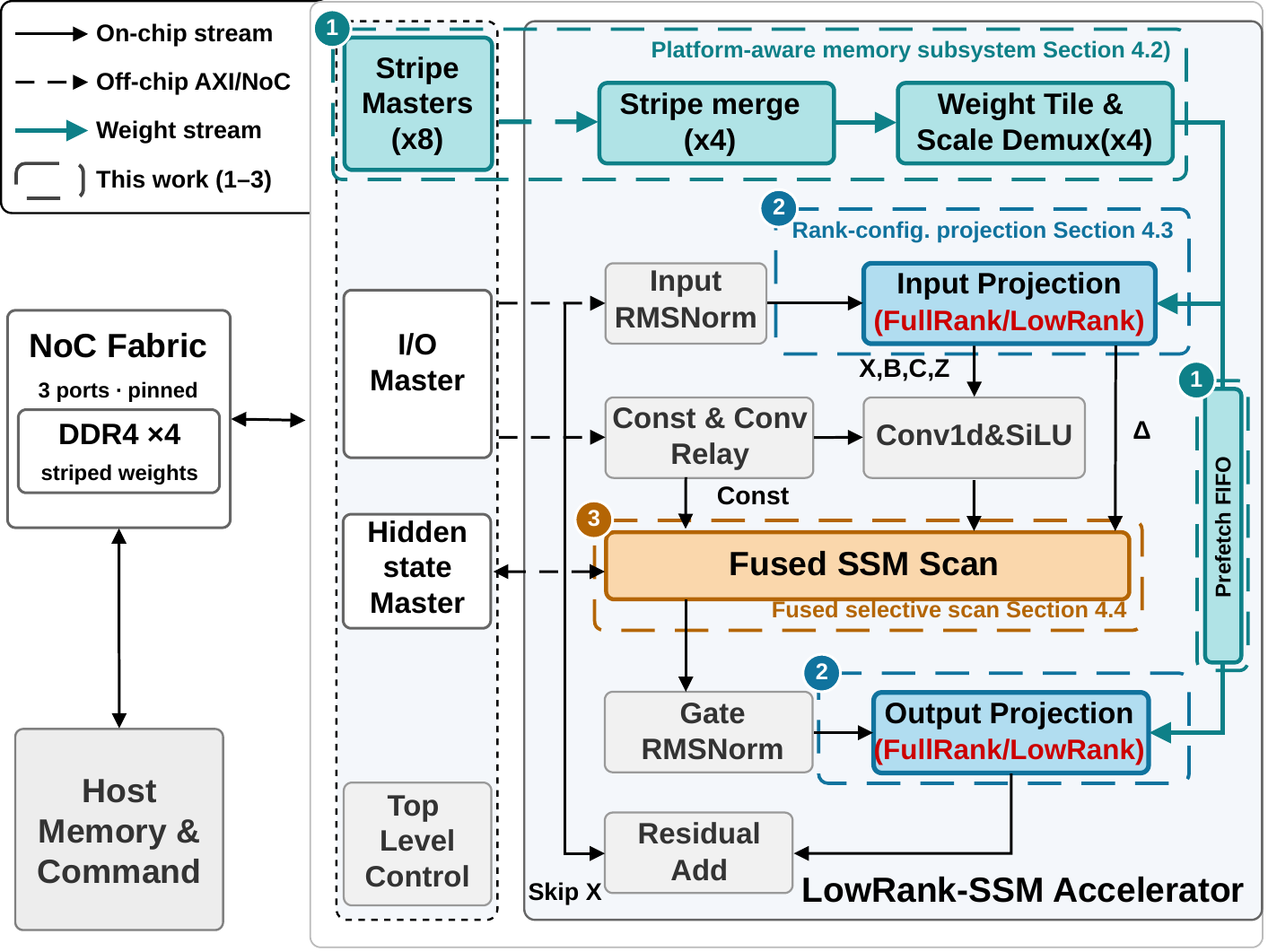}
    \caption{Overall architecture of the LowRank-SSM kernel.} 
    \label{fig:arch}
\end{figure}

\subsection{Overall Hardware Architecture}
\label{sec:hw:baseline}
As shown in Figure~\ref{fig:arch}, LowRank-SSM processes the $L$ layers of the model in sequence on one dataflow kernel. Each layer passes through the Input RMSNorm, the Input Projection, the Conv1d\&SiLU, the Fused SSM Scan, the Gate RMSNorm and the Output Projection, and the Residual Add adds the result to the skip path. The modules are connected by on-chip stream FIFOs, so each module starts as soon as its first operand arrives, and the token stays on chip between layers. The grey blocks are conventional; the three coloured groups are where this work departs from prior SSM accelerators, and the rest of this section is organized around them.

\textbf{Memory subsystem (teal).} In single-token decode every projection weight is used once, so the kernel is bound by how fast weights arrive, and the memory path is designed first. The VC1902 exposes its four DDR4 controllers to a kernel through three NoC ports, and the weight image in DDR4 is packed per weight stream and striped over two AXI masters. Eight Stripe Masters, two per stream, read it with their NoC placement balanced over the three ports and pinned across builds; four Stripe Merge units reassemble the four streams, and four Weight Tile \& Scale Demux units separate the weight tiles from the group scales embedded in the same stream. On the path to the output projection, deep UltraRAM Prefetch FIFOs hold the output-projection rows that the streams keep fetching while the input projection drains and the scan runs. Section~\ref{sec:memsys} shows that these choices, not burst tuning, decide the delivered bandwidth.
 
\textbf{Projection units (blue).} The Input Projection and the Output Projection run every layer in full-rank or low-rank mode. In Section~\ref{sec:projunit}, the rank of each layer comes from a per-layer rank table and enters the units as a loop bound, so both factor stages of a low-rank layer run on the same GEMV engines and streams as a dense layer, and a new rank schedule changes the weight image and the rank table but not the bitstream.
 
\textbf{Fused scan (orange).} In Section~\ref{sec:scan}, the Fused SSM Scan evaluates the state update, the output contraction and the gating in one pipelined pass, with the recurrent state streamed through the Hidden-state Master rather than held on chip.
 
The remaining masters carry little traffic: the I/O Master moves the token and the per-layer data, the Const \& Conv Relay forwards the convolution weights and scan constants of each layer and fetches those of the next layer while the current one runs, and the Top Level Control sequences the layers.

\subsection{Platform-Aware Memory Subsystem}
\label{sec:memsys}
\textbf{Bandwidth model.} In the decode stage, each projection weight is used once per token, so the projections are bounded by the off-chip bandwidth. On NoC-based FPGAs such as AMD Versal, a kernel reaches the memory controllers through a small number $P$ of NoC ports, each provisioned with a bandwidth $B_{\mathrm{port}}$, and each AXI master enters the NoC through one port. A single master saturates at a bandwidth $b_m$, independent of its burst length. The bandwidth delivered to the kernel is therefore
\begin{equation}
BW_{\mathrm{eff}} = \sum_{p=1}^{P} \min\big(n_p\, b_m,\ \eta\,B_{\mathrm{port}}\big)
\;\le\; \min\big(\eta\,P\,B_{\mathrm{port}},\ B_{\mathrm{DDR}}\big),
\label{eq:bwport}
\end{equation}
where $n_p$ is the number of active masters on port $p$, $\eta<1$ the efficiency of the NoC and the DRAM under load, and $B_{\mathrm{DDR}}$ the peak bandwidth of the memory controllers. On our platform, the kernel reaches the memory through $P=3$ NoC ports of $B_{\mathrm{port}}\approx17.27$\,GB/s each, a single master saturates at $b_m\approx12.55$\,GB/s, and $\eta$ is close to 0.8 under the full load of the kernel. Each port adds the bandwidth of its masters until it saturates, and on this platform 2 masters already fill a port. The delivered bandwidth thus grows with the number of masters and with how evenly they are spread over the ports, and the three ports, whose sum is about half of $B_{\mathrm{DDR}}$, rather than the DRAM devices set the ceiling.

\textbf{Bandwidth design.} Following Eq.~\eqref{eq:bwport}, we use more masters, spread them evenly over the ports, and keep them busy. 1)~The weights that an engine reads in a layer are packed on the host into one stream, in reading order and with the group scales embedded, and the stream is striped in burst-sized chunks over 2 Stripe Masters, so each master reads one linear sweep per layer. 2)~The 2 stripes of a stream are consumed in lock-step, so a stream advances at the pace of its slower master. We therefore spread the masters evenly over the ports and run the output projection on three streams whose six masters occupy the three ports two by two. 3)~Each output-projection stream prefetches its rows into a deep on-chip FIFO while the input projection drains and the scan runs. 4)~We pin the NoC placement of the masters to a reference solution, which makes the delivered bandwidth reproducible across builds. With this memory subsystem, a projection takes as long as its bytes take to arrive, provided that each engine receives a share of the bytes in proportion to the bandwidth of its stream. The projection unit below is designed around this condition.

\begin{figure*}[t]
    \centering
    \includegraphics[width=\textwidth]{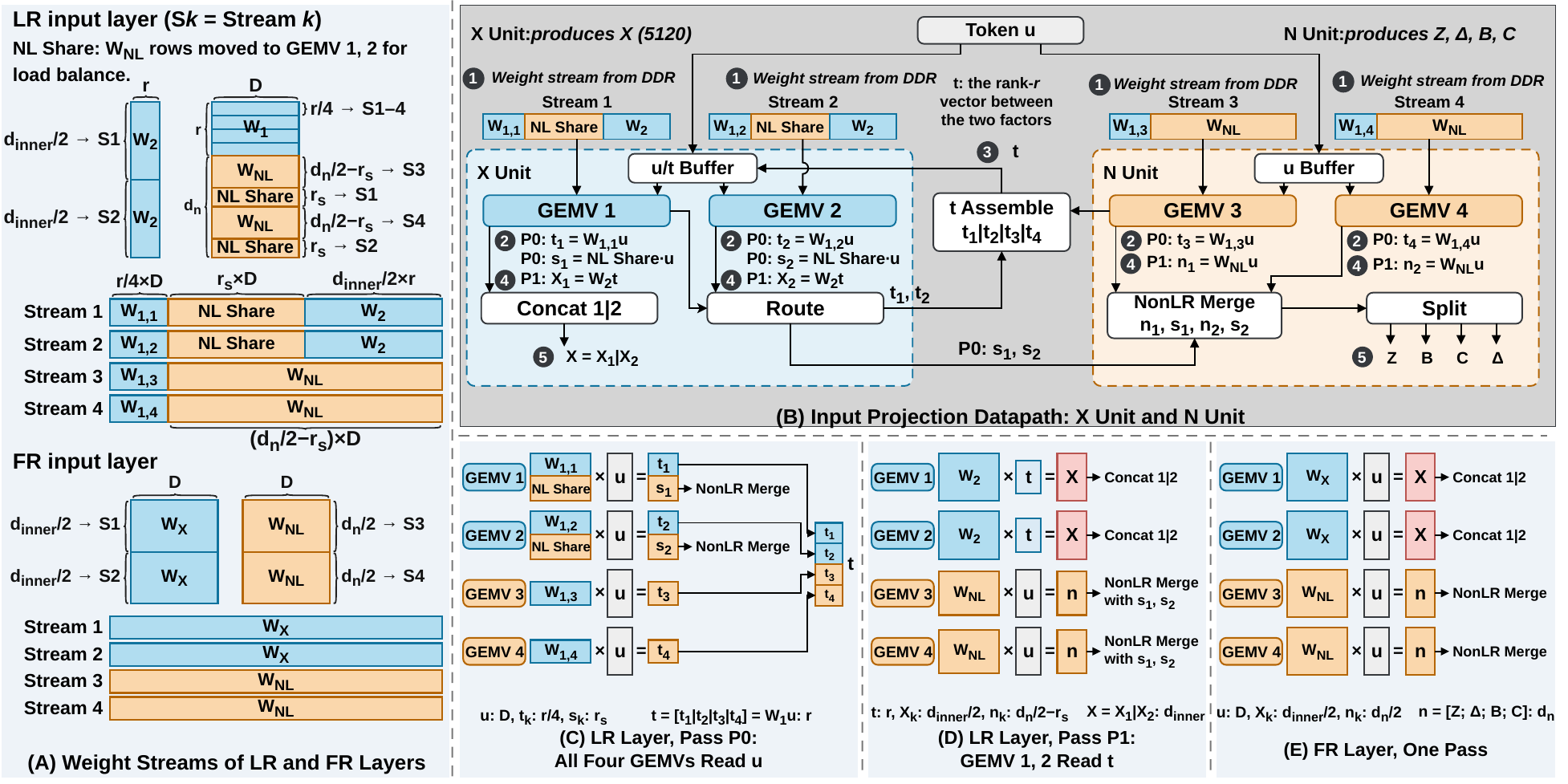}
    \caption{Input projection. (A) Content of the four weight streams in an LR and an FR layer. (B) Datapath of the X Unit and the N Unit. (C)--(E) Work of each GEMV engine in the two passes of an LR layer and in the single pass of an FR layer.} 
    \label{fig:INPUT_PROJ}
\end{figure*}

\subsection{Rank-Configurable Projection Unit}
\label{sec:projunit}

\textbf{Overview.} A projection is low-rank (LR) when its weight matrix is replaced by two factors of rank $r$, and full-rank (FR) otherwise. For the matrix $W_X$ that produces $X$, the factors $W_2\in\mathbb{R}^{d_{\mathrm{inner}}\times r}$ and $W_1\in\mathbb{R}^{r\times D}$ change the weight traffic by
\begin{equation}
\rho(r) = \frac{r\,(D+d_{\mathrm{inner}})}{D\,d_{\mathrm{inner}}},
\label{eq:traffic}
\end{equation}
where the numerator counts the weights of the two factors and the denominator those of the dense matrix, so an LR layer reads fewer bytes whenever $\rho(r)<1$. As shown in Figure~\ref{fig:INPUT_PROJ}(B), the input projection includes two units with two GEMV engines each, and both read the normalized token $u$, which holds $D=2560$ values in Mamba2-2.7B. The X Unit, with GEMV~1 and 2 and an u/t Buffer, produces $X$ with $d_{\mathrm{inner}}=5120$ values. The N Unit, with GEMV~3 and 4 and a u Buffer, produces the other operands $n=[Z;\Delta;B;C]$ from the rows $W_{\mathrm{NL}}$, which stay dense in every layer: $Z$ has another $d_{\mathrm{inner}}$ values, $\Delta$ one value per head, 80 in total, and $B$ and $C$ have $N=128$ values each, so $n$ holds 5456 values. Each engine reads one stream: GEMV~$k$ reads Stream~$k$. 
The key hardware property is that low rank introduces no dedicated low-rank engine, stream, or FIFO. Both factors reuse the four existing input-projection streams and GEMV engines, while the intermediate vector remains on chip. Rank therefore changes the amount of data processed by the same datapath rather than selecting a different hardware specialization.

\textbf{Full-rank layer.} In an FR layer, Streams~1 and 2 carry the two halves of $W_X$, and Streams~3 and 4 carry the two halves of $W_{\mathrm{NL}}$, as listed in Figure~\ref{fig:INPUT_PROJ}(A). The projection runs in one pass, shown in Figure~\ref{fig:INPUT_PROJ}(E): all four engines multiply their rows by the token $u$. Concat~1$|$2 joins the two halves of $X$, 2560 values each, and the Split unit sends $Z$, $B$, $C$ and $\Delta$ to the next modules.

\textbf{Low-rank layer.} An LR layer runs in two passes, in the five steps \ding{182}--\ding{186} of Figure~\ref{fig:INPUT_PROJ}(B). \ding{182}~Each Stream~$k$ starts with $W_{1,k}$, the $k$-th quarter of the rows of $W_1$, and runs straight into GEMV~$k$. \ding{183}~In pass P0, shown in Figure~\ref{fig:INPUT_PROJ}(C), all four engines read $u$, and GEMV~$k$ computes $t_k=W_{1,k}\,u$, a quarter of the intermediate vector and $r/4$ values long. \ding{184}~The t Assemble unit collects $t_1$ to $t_4$ into $t=W_1u$, $r$ values in all, and writes it into the u/t Buffer; the deployed schedule sets $r$ to 896, 768 or 640, so $t$ never exceeds 896 of the 2560 values that $u$ holds. \ding{185}~In pass P1, shown in Figure~\ref{fig:INPUT_PROJ}(D), GEMV~1 and 2 read $t$ and produce the halves $X_1$ and $X_2$ of $X=W_2\,t$, 2560 values each, while GEMV~3 and 4 continue with $W_{\mathrm{NL}}\,u$ and produce $n_1$ and $n_2$. \ding{186}~Concat~1$|$2 joins $X_1$ and $X_2$ into $X$, and the NonLR Merge rebuilds $n$ before the Split unit sends $Z$, $B$, $C$ and $\Delta$ to the next modules. Every row of $W_2$ thus meets the complete vector $t$ inside one engine, so the cascade needs a gather of $r$ values and no sum of partial results across engines.

\textbf{Balancing the two units.} Following the bandwidth design, the four engines should stream the same number of bytes. In an LR layer, however, the X Unit streams short rows of $r$ columns, whereas the N Unit streams $W_{\mathrm{NL}}$ rows of $D$ columns regardless of the rank, so the N Unit would finish last. We therefore move $m$ rows of $W_{\mathrm{NL}}$ per engine to the X Unit, shown as NL Share in Figure~\ref{fig:INPUT_PROJ}(A). GEMV~1 and 2 compute these rows in P0, next to $t_1$ and $t_2$; the Route unit forwards their results $s_1$ and $s_2$, and the NonLR Merge returns them to $n$ in the original row order. Equating the rows streamed by an engine of each unit, counted in rows of $D$ columns, gives
\begin{equation}
m = \tfrac{1}{2}\Big(n_h - \tfrac{d_{\mathrm{inner}}}{2D}\,r\Big),
\label{eq:share}
\end{equation}
where $n_h=|n|/2$ is the number of $W_{\mathrm{NL}}$ rows per N-Unit engine and $\tfrac{d_{\mathrm{inner}}}{2D}\,r$ counts the $W_2$ rows of an X-Unit engine in rows of $D$ columns. A smaller rank moves more rows: $m$ is 912, 976 and 1040 for the three deployed ranks, so each N-Unit engine keeps 1816, 1752 or 1688 rows and the four parts $n_1$, $s_1$, $n_2$ and $s_2$ still add up to the 5456 values of $n$. We round $m$ down to a multiple of the output vector width, and FR layers keep $m=0$.

\textbf{Output projection.} The output projection follows the same scheme on three engines, fed by the three port-balanced streams. In an LR layer, the rows of both factors are divided evenly over the three engines and appended to the tails of their streams, and the two passes produce the intermediate vector, 1152 values at the deployed rank, and then the 2560 values of the layer output. These rows are prefetched while the input projection drains and the scan runs.

\begin{figure}[t]
    \centering
    \includegraphics[width=\linewidth]{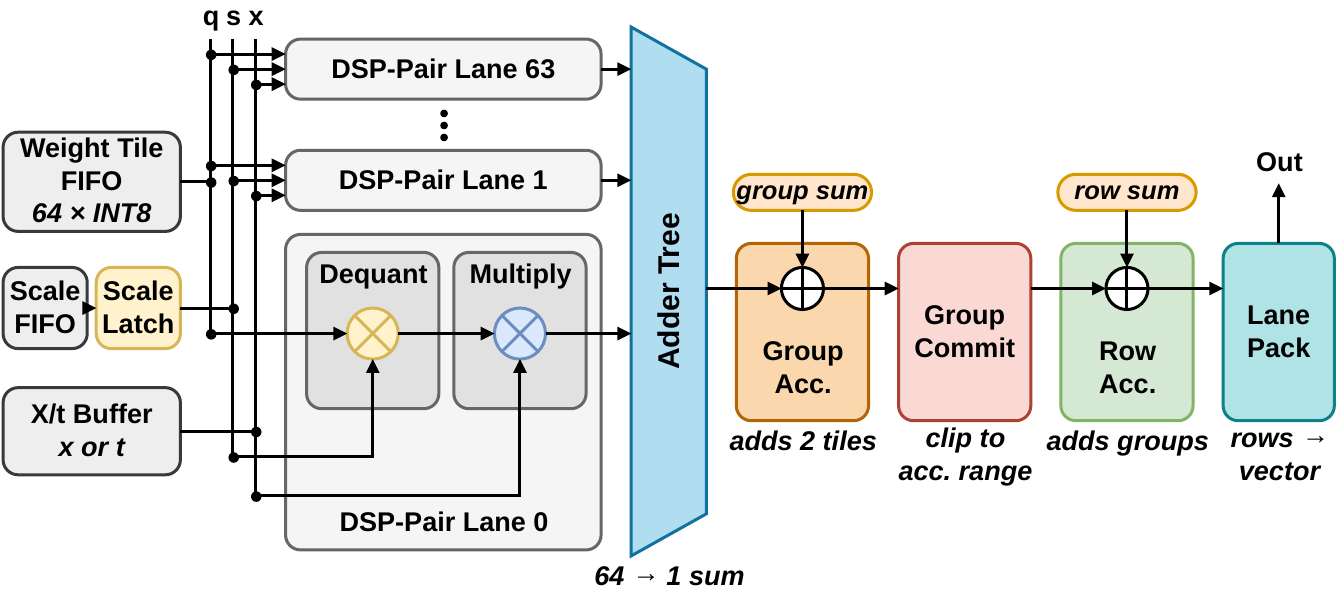}
    \caption{Quantized GEMV engine of the input projection.} 
    \label{fig:QGEMV}
\end{figure}

\textbf{Quantized GEMV engine.}\label{sec:gemv} All projection engines share one quantized GEMV engine. As shown in Figure~\ref{fig:QGEMV}, it consumes one weight tile per cycle, which holds $\beta=64$ INT8 weights and is exactly one 512-bit memory beat. The Weight Tile FIFO, the Scale Latch and the u/t Buffer drive three buses that every lane taps: $q$ carries the 8-bit weight code, $s$ the 24-bit Q8.16 group scale, and $x$ the activation, Q6.10 in the input projection and Q12.20 in the output projection. We write Q$a$.$b$ for a signed fixed-point word with $a$ integer bits and $b$ fraction bits, so Q8.16 is 24 bits wide, Q6.10 is 16 and Q12.20 is 32. The output projection takes the wider activation because it carries the residual stream, which grows across the 64 layers. One scale serves a group of $g=128$ columns. Therefore the Scale Latch holds it across two consecutive tiles, and the stream sends it once ahead of a block of rows instead of repeating it per tile. Each lane is a DSP pair, drawn open for Lane~0. The Dequant DSP multiplies $q$ by $s$, and the Multiply DSP multiplies the result by $x$. An input-projection engine therefore costs 128 DSPs, and an output-projection engine 193, because its wider Q12.20 activations need more. The Adder Tree reduces the $\beta$ products to one partial sum. Accumulating a whole row directly in the Q12.20 accumulator would round away the low bits, so the engine accumulates in two levels. The Group Acc. adds each partial sum into a 48-bit group sum, which is wide enough to hold one group without loss. The Group Commit then clips the finished group sum to the accumulator range, once per group and so once every two tiles, and the Row Acc. adds it to the row sum. When a row ends, the Lane Pack writes the row sum into an output vector. The row length is a run-time value: $D=2560$ for the rows of $W_X$, $W_{\mathrm{NL}}$ and $W_1$, and $r$ for the rows of $W_2$, which the deployed schedule sets to 640, 768 or 896 in the input projection and 1152 in the output projection. In this way, one engine serves FR and LR layers and both passes without a pipeline restart.

\subsection{Fused Selective Scan Unit}
\label{sec:scan}

The fused scan unit evaluates the selective state update of Mamba-2 for one token. It receives $x_t$ from the Conv1d\&SiLU, $Z$, $B$, $C$ and $\Delta$ from the input projection, and the constants $A$ and $D_{\mathrm{skip}}$ from the Const \& Conv Relay. For each head, it first computes the step size and the decay,
\begin{equation}
\delta=\mathrm{softplus}(\Delta),\qquad \bar{A}=\exp(\delta A),
\label{eq:disc}
\end{equation}
and then updates the state $h\in\mathbb{R}^{N\times d_{\mathrm{inner}}}$ and forms the gated output,
\begin{align}
h_t &= \bar{A}\,h_{t-1} + (\delta B)\,x_t, \nonumber\\
\hat{y}_t &= \big(C^{\top}h_t + D_{\mathrm{skip}}\odot x_t\big)\odot\mathrm{SiLU}(Z).
\label{eq:gate}
\end{align}
The unit walks over the state in $V\times V$ tiles of $8\times8$ elements, one tile per cycle, so the $128\times5120$ state of a Mamba2-2.7B layer passes through the unit as $16\times640$ tiles. Within a tile, $\delta B$ is computed once per state lane and shared by the $V$ channel lanes, each lane updates its element of $h$, and the products with $C$ are accumulated across the state tiles into one entry per channel. A final pass over the $d_{\mathrm{inner}}$ channels adds the skip term, applies the gate, and hands $\hat{y}_t$ to the Gate RMSNorm.

%% file: sections/05-Experiments.tex
\section{Experiments}
\label{sec:experiments}

\begin{table*}[t]
\centering
\caption{Downstream task accuracy for Mamba2-2.7B under low-rank
  compression and quantization. PPL is reported for LAMBADA and WikiText. LR represents low rank.}
\label{tab:mamba2_bench}
\setlength{\tabcolsep}{5pt}
\renewcommand{\arraystretch}{1}
\resizebox{0.88\textwidth}{!}{%
\begin{tabular}{l | c c c c c c c c c}
\toprule
\textbf{Method}
  & \textbf{LAM PPL\,$\downarrow$}
  & \textbf{WikiText PPL\,$\downarrow$}
  & \textbf{LAM\,$\uparrow$}
  & \textbf{Hella\,$\uparrow$}
  & \textbf{ARC-E\,$\uparrow$}
  & \textbf{ARC-C\,$\uparrow$}
  & \textbf{Wino\,$\uparrow$}
  & \textbf{OBQA\,$\uparrow$}
  & \textbf{Avg\,$\uparrow$} \\
\midrule
FP16 (Dense) \textit{Baseline}
  & 4.09 & 11.46 & 69.50 & 66.60 & 69.50 & 36.30 & 63.90 & 38.80 & 57.49 \\
LowRank-SSM (LR only)
  & 4.59 & 12.05 & 67.38 & 63.30 & 67.72 & 35.24 & 63.22 & 38.20 & 55.84 \\
\midrule
\multicolumn{10}{l}{\textit{INT8 quantization}} \\
LightMamba INT8
  & \textbf{4.07} & -- & \textbf{69.70} & 66.50 & 69.30 & \textbf{36.90} & \textbf{64.00} & 38.80 & 57.50 \\
LowRank-SSM (INT8 only)
  & 4.09 & \textbf{11.49} & 69.50 & \textbf{66.50} & \textbf{69.60} & 36.40 & 63.90 & 38.80 & 57.50 \\
LowRank-SSM (LR + INT8)
  & 4.59 & 12.05 & 67.58 & 63.40 & 67.62 & 35.04 & 63.62 & 38.20 & 55.91 \\
\midrule
\multicolumn{10}{l}{\textit{INT4 quantization}} \\
LightMamba INT4
  & 6.48 & -- & 57.30 & 62.70 & 65.50 & 35.30 & 60.70 & 37.60 & 56.30 \\
LowRank-SSM (INT4 only)
  & \textbf{4.44} & \textbf{12.07} & \textbf{67.40} & \textbf{65.90} & \textbf{68.90} & \textbf{36.60} & 62.90 & \textbf{38.20} & 56.60 \\
LowRank-SSM (LR + INT4)
  & 5.21 & 12.66 & 64.38 & 62.50 & 66.02 & 34.44 & \textbf{62.92} & 37.60 & 54.64 \\
\bottomrule
\end{tabular}%
}
\end{table*}

\subsection{Experimental Setup}
 \label{sec:exp_setup}

\textbf{Model and software.}  The hardware evaluation and the ablation study use Mamba2-2.7B~\cite{Dao2024}.
All weights are stored as INT8 with one symmetric scale per output row and 128-column input group.
Low-rank factors are pre-computed offline by whitened truncated SVD (SVD-LLM~\cite{wang2025svdllmtruncationawaresingularvalue}) of the floating-point weights,
$W \approx W_2W_1$ with $W_1=\sqrt{\Sigma_r}V_r^{\top}$ and $W_2=U_r\sqrt{\Sigma_r}$, and are then
quantized to INT8 in the same group-wise format as the dense weights; no gradient-based fine-tuning is
performed at any stage.  Every hardware variant is verified bit-exactly against its own fixed-point
reference. Accuracy is evaluated on six zero-shot benchmarks:
LAMBADA~\cite{paperno2016lambadadatasetwordprediction}, HellaSwag~\cite{hellaswag},
ARC-Easy and ARC-Challenge~\cite{arc}, Winogrande~\cite{winogrande}, and OpenbookQA~\cite{obqa}, together
with LAMBADA~\cite{paperno2016lambadadatasetwordprediction} perplexity and WikiText-2~\cite{wikitext} perplexity at a sequence length of 2048.

\textbf{Rank schedule.} LowRank-SSM uses one per-projection rank schedule, found by the search of Section~3.2 with a 5\% ceiling on WikiText-2 perplexity and marked in Figure~3. It compresses 25 of the 128 projections, saves 7.74\% of the target weights, which is 5.0\% of the weights read per token. Selected ranks are rounded to multiples of 128 for FPGA deployment.

\begin{figure}[t]
  \centering
  \includegraphics[width=\columnwidth]{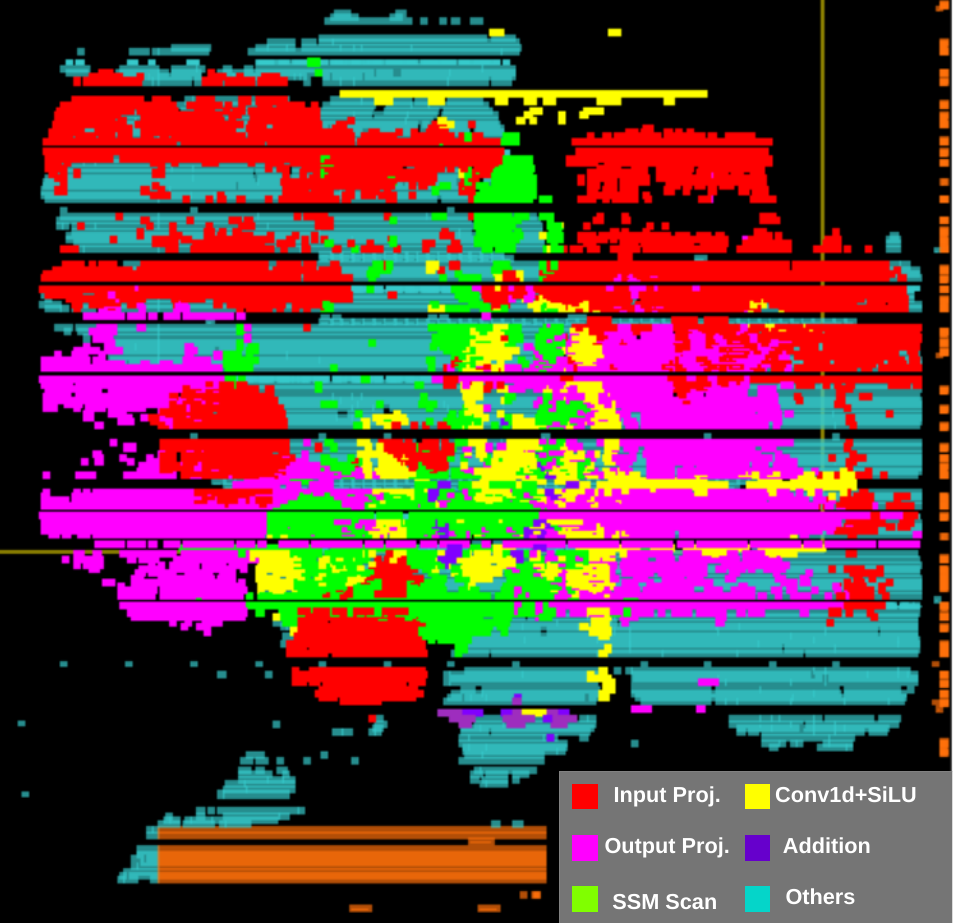}
  \caption{LowRank-SSM implementation layout on the VC1902.}
  \label{fig:layout}
\end{figure}

\textbf{Hardware and quantization.}  All projection weights, including the low-rank factors, are quantized
to INT8 with per-row, per-group scaling ($g{=}128$). Activations are represented as \texttt{ap\_fixed<16,6>}
and intermediate accumulations as \texttt{ap\_fixed<32,12>}. The kernel is synthesized with Vitis~HLS~2022.1 and implemented with Vitis~2022.1 at 230~MHz on the AMD Versal
VC1902 FPGA, on which all measurements are taken. Figure~\ref{fig:layout} shows the layout of LowRank-SSM on the
VC1902 FPGA.

\textbf{Baselines.}  We compare against: (i)~FP16 dense baselines;(ii)~the FPGA accelerators LightMamba~\cite{wei2025} and FastMamba~\cite{wang2025fastmambahighspeedefficientmamba}, both with INT8 weights on Mamba2-2.7B. FastMamba is re-implemented on the same VC1902 and clocked at the same 230~MHz; (iii)~low-rank only (no quantization) at $\Delta=5\%$~PPL budgets to isolate the contribution of rank reduction. The mixed ranks were searched by our per-projection greedy search in Section~\ref{sec:method:greedy} with decomposition from SVD-LLM~\cite{wang2025svdllmtruncationawaresingularvalue}; (iv)~weight-only quantization in Section~\ref{sec:method:quantization} with 4 bit and 8 bit;and (v)~for deployment context, an RTX~4090 GPU on the same host running Mamba2-2.7B with both full-rank and low-rank projections with torch CUDA graph.

\subsection{Accuracy Evaluation}
\label{sec:exp:accuracy}
 
\textbf{Multi-Task Benchmark.} Table~\ref{tab:mamba2_bench} lists the perplexities and accuracy of all configurations on the six tasks.

\textbf{Quantization only.} With per-row, per-group INT8 quantization and no rotation, LowRank-SSM INT8 obtains a LAMBADA perplexity of 4.09 and an average accuracy of 57.5\%, the same as the FP16 baseline and close to LightMamba INT8 (4.07, 57.5\%). With INT4 weights at full rank, it obtains 4.44 and 56.6\%, compared with 6.48 and 56.3\% for LightMamba INT4. On these projection weights, per-group INT4 quantization without rotation gives a lower LAMBADA perplexity than the rotation-based scheme.


\textbf{Low-rank only.} At the deployed 5\% ceiling, rank reduction alone lowers the average accuracy by 1.6\,pp, from 57.4\% to 55.8\%, and raises the LAMBADA perplexity from 4.09 to 4.59 and the WikiText-2 perplexity from 11.46 to 12.05. At a 20\% ceiling , which is not shown in the table, the decrease is 5.4\,pp, to 52.0\%, and LAMBADA accuracy falls by 10.5\,pp, which is why LowRank-SSM deploys the 5\% schedule. 

\textbf{Combining rank reduction and quantization.} INT8 adds nothing to the cost of rank reduction. LowRank-SSM (LR + INT8) reaches 55.9\% average accuracy and a LAMBADA perplexity of 4.59, the same as the low-rank model in FP16 (55.8\%, 4.59), so the whole 1.5\,pp gap to the dense model comes from the rank schedule. At INT4 the two errors compound slightly: INT4 alone costs 0.8\,pp and rank reduction alone 1.6\,pp, while LR + INT4 loses 2.8\,pp and its LAMBADA perplexity rises to 5.21. Rank reduction acts on the singular-value structure of the weights and quantization on their numeric representation, so at INT8 the rank schedule can be chosen without regard to the quantizer.

\textbf{Comparison with INT4.} Four-bit weights are the other way to reduce weight traffic, and they stack with rank reduction rather than compete with it. LightMamba INT4, which also quantizes the activations to four bits, halves the traffic at 57.3\% LAMBADA accuracy and a LAMBADA perplexity of 6.48. Our weight-only INT4 halves it at 67.4\% and 4.44, and applying the low rank on top removes a further 5.0\% of the bytes at 64.38\% and 5.21, still ahead of LightMamba INT4. LowRank-SSM deploys INT8 because it is lossless on these tasks, which leaves rank as the only lever on traffic at that precision; LR + INT8 retains 55.9\% and 4.59.

\subsection{Hardware Evaluation}
\label{sec:exp:hw}

\begin{table}[t]
\centering
\caption{Post-route resource utilization on the VC1902.}
\label{tab:hw_summary}
\begin{tabular}{lrr}
\toprule
\textbf{Resource} & \textbf{Used} & \textbf{Util.\,\%} \\
\midrule
LUT   & 233{,}502 & 25.95 \\
FF    & 237{,}647 & 13.20 \\
DSP   &   1{,}624 & 82.52 \\
BRAM  &     309.5 & 32.01 \\
URAM  &       101 & 21.81 \\
\bottomrule
\end{tabular}
\vspace{-5 mm}
\end{table}

Table~\ref{tab:hw_summary} reports the post-route utilization of the 64-layer
mixed-rank design on the Versal VC1902, including the platform shell,
implemented at 230\,MHz with a worst negative slack of $+0.091$\,ns. DSP is the most used resource at
1{,}624 of 1{,}968 slices, or 82.52\%. The seven quantized GEMV engines
account for most of it: each lane of an engine is a
DSP pair that dequantizes one INT8 weight and applies it to the activation, so
each of the four input-projection engines occupies 128 DSP slices to retire one
64-weight beat per cycle, and each of the three output-projection engines,
whose activations stay in Q12.20, occupies 193; the fused scan adds its
$V\!\times\!V$ multiplier array. The two factor stages of the low-rank
projections are computed by these same engines, so no multipliers are spent on
dedicated low-rank engines.
BRAM and URAM, at 32.01\% and 21.81\%, hold the tile and scale FIFOs of the
weight streams, the 8192-tile UltraRAM FIFOs in which the output-projection
rows are prefetched, and the on-chip layer buffers; because the fused scan
contracts every state tile in the iteration that produces it, no memory scales
with $N d_{\mathrm{inner}}$. LUT and FF utilization remain moderate at 25.95\%
and 13.20\%.

\begin{table}[b]
  \centering
  \caption{Comparison on Mamba2-2.7B.}
  \label{tab:comparison}
  \setlength{\tabcolsep}{3.5pt}
  \renewcommand{\arraystretch}{1.1}
  \begin{tabular}{@{}lcccc@{}}
    \toprule
     & \textbf{Device} & \textbf{Thpt.} & \textbf{Power} & \textbf{Energy Eff.} \\
     &                 & \textbf{(tok/s)} & \textbf{(W)} & \textbf{(tok/s/W)} \\
    \midrule
      Full-rank FP16 & RTX 4090 & 145.61 & 292.15 & 0.498 \\
      Low-rank FP16 & RTX 4090 & 147.86 & 287.67 & 0.514 \\
      FastMamba~\cite{wang2025fastmambahighspeedefficientmamba} & VC1902 & 5.307 & 3.00 & 1.77 \\
    LightMamba~\cite{wei2025}                                 & VC1902 & 3.61  & 2.49 & 1.45 \\
    \textbf{LowRank-SSM}                                      & VC1902 & \textbf{16.79} & 6.90 & \textbf{2.43} \\
  
    \bottomrule
  \end{tabular}
\end{table}

Table~\ref{tab:comparison} compares LowRank-SSM with prior FPGA-based SSM
accelerators and with a GPU. All workloads are Mamba2-2.7B with 64 layers and
single-token decode, and the FPGA designs use INT8 weights. LowRank-SSM and FastMamba were measured on the same Versal VC1902 board with the same flow:
throughput is the mean over repeated 64-layer single-token decode launches, and power is
the XRT-reported board power (\texttt{power\_consumption\_watts}) during those launches minus the XRT-reported idle power of the same board with the design's bitstream loaded.
The LightMamba numbers are the published W8A8 result at 400\,MHz on a VCK190, whose memory path provides about 12\,GB/s~\cite{wei2025}. Every FPGA design we measured is bit-exact against its own reference on the board.
SpecMamba~\cite{zhong2025specmambaacceleratingmambainference} reports 20.6 tok/s on the same device with INT4 weights and speculative decoding; both are orthogonal to the memory path and the rank-configurable datapath presented here.


\textbf{FPGA comparison.} LowRank-SSM, implemented at 230\,MHz, reaches 16.79 tok/s, or 59.6\,ms per token, at 6.9\,W (2.43 tok/s/W). LightMamba~\cite{wei2025} reports 3.61 tok/s at 2.49\,W (1.45 tok/s/W); LowRank-SSM is 4.65$\times$ faster and 1.68$\times$ more energy-efficient. FastMamba~\cite{wang2025fastmambahighspeedefficientmamba} originally targets
the Xilinx VC709, whose DSP density and memory bandwidth differ from the
VC1902's, so its published numbers are not directly comparable. We re-implement
its architecture on the VC1902 with its original dataflow and quantization scheme;
it reaches 5.307~tok/s at 3.00\,W (1.77~tok/s/W). LowRank-SSM is 3.16$\times$ faster and 1.38$\times$ more energy-efficient. 

\textbf{GPU baseline.} On an RTX 4090 on the same host, Mamba2-2.7B runs at 145.61 tok/s and 292.15\,W (0.498~tok/s/W) with full-rank projections and at 147.86 tok/s and 287.67\,W (0.514~tok/s/W) with the low-rank projections. The GPU is 8.8$\times$ faster than LowRank-SSM, and LowRank-SSM is 4.73$\times$ more energy-efficient than the low-rank GPU run. With CUDA graphs the low-rank model is 1.5\% faster than dense for 5.0\% fewer bytes, so the GPU converts about 30\% of the removed traffic into time; the FPGA converts 93\%.

\textbf{Memory path and port balance.} A read-only probe on the board fixes the constants of Eq.~(3): one AXI master saturates at 12.55\,GB/s, and six masters at two per port deliver 41.45\,GB/s, 80\% of the provisioned 51.8\,GB/s ($\eta \approx 0.80$). The projection engines can absorb 58.9\,GB/s (input) and 44.2\,GB/s (output), so both projection phases are memory-bound. Placement matters more than the number of masters: moving the same output-projection bytes, six masters placed 2/2/2 on the three ports are 1.49$\times$ faster than four placed 2/1/1, whereas six placed 3/1/2 are slower (0.91$\times$) and eight placed 3/2/3 reach only 1.21$\times$. The final design therefore runs the output projection on three streams placed 2/2/2.

\textbf{Delivered bandwidth and rank balance.} Averaged over a token, the design sustains about 42\,GB/s, 81\% of the provisioned bandwidth and in line with the six-master measurement. We examine the balance per layer type with the more aggressive schedule of Section~6, which uses three input ranks. Every layer type reaches between 40.7 and 42.7\,GB/s, so low-rank layers stream at the same rate as dense ones, which is the condition under which saved bytes become saved time. Eq.~(5) moves $m = 912$, 976 and 1040 rows for ranks 896, 768 and 640; each engine of a rank-896 layer then streams 81,280 to 81,600 beats, against 109,120 on the critical engine of a dense layer. Over a whole token, 93\% of the traffic removed by the low-rank projections appears as latency.

\section{Ablation Study}
\label{sec:ablation}
\begin{table}[t]
  \centering
  \caption{Ablation on the VC1902 for Mamba2-2.7B, 64-layer single-token decode at 230\,MHz. Each row adds one component of the co-design stack to the row above. The LR and FS rows use the ablation schedule (WikiText-2 perplexity 13.24))}
  \label{tab:ablation}
  \footnotesize
  \setlength{\tabcolsep}{2.2pt}
  \begin{tabular}{@{}lrrrrrrrrr@{}}
    \toprule
            & Latency & Tok./ & Speed & Traffic   & LUT & FF  & DSP & BRAM & URAM \\
    Variant & (ms)    & s     & Up & (GB)    & (k) & (k) &     &      &      \\
    \midrule
    Baseline    & 234.9 &  4.257 & 1$\times$ & 5.14  & 162 & 174 & 1179 & 311   & 226 \\
    + SP        & 117.1 &  8.536 & 2.01$\times$ & 5.14  & 205 & 201 & 1120 & 308   & 265 \\
    + Q         &  65.4 & 15.299 & 1.79$\times$ & 2.64 & 229 & 234 & 1589 & 309.5 & 223 \\
    + LR        &  56.9 & 17.566 & 1.15$\times$ & 2.27  & 235 & 239 & 1624 & 309.5 & 223 \\
    + FS        & \textbf{54.3} & \textbf{18.43} & 1.05$\times$ & 2.27  & 233 & 237 & 1624 & 309.5 & \textbf{101} \\
    \bottomrule
  \end{tabular}
\end{table}
The ablation, Table~\ref{tab:ablation}, is designed to test the central co-design hypothesis: structural rank reduction is effective only when the memory system and projection datapath expose its reduced traffic to the execution time. We therefore add the four techniques incrementally while holding the final implementation structure and NoC placement fixed. In particular, the LR step isolates the benefit of the rank allocation and rank-configurable datapath after the memory subsystem has already been optimized. We build five bitstreams that differ by exactly one technique each, in the order in which the techniques were added. The LR and FS rows use more aggressive rank schedule than the deployed one, with a WikiText-2 perplexity of 13.24 against 12.05, which removes 13.9\% of the projection traffic against 5.0\%. Because rank is a run-time table, the deployed schedule runs on the same final bitstream without a rebuild.

Every variant is bit-exact against the fixed-point reference, and latency, traffic, and bandwidth are measured. The complete co-designed system is $4.33\times$ faster than the baseline. The LR step itself reduces latency from 65.4 ms to 56.9 ms (12.9\%) by reducing projection traffic, demonstrating that the rank-configurable datapath converts the software compression into measured execution-time
savings. Low-rank compression becomes useful on FPGA only after the memory subsystem and datapath are co-designed to convert reduced weight traffic into latency reduction.

The steps, in the order they are added, are as follows. \textbf{Striped packed streaming (SP)}, Section~\ref{sec:memsys}: each engine reads one packed stream striped over two AXI masters, and the output projection runs on three engines whose six masters sit two per NoC port, with its rows prefetched on chip. The baseline instead feeds each projection half from a single master and has two output-projection engines. \textbf{Group-wise INT8 (Q)}, Section~\ref{sec:method:quantization}, replaces 16-bit fixed-point weights. \textbf{Low-rank projections (LR)}, Section~\ref{sec:projunit}: selected layers compute $W_2 W_1 x$ on the engines and streams of the dense projection. \textbf{Fused scan (FS)}, Section~\ref{sec:scan}: one pipelined pass replaces a three-pass scan.

In Table~\ref{tab:ablation}, the latency follows the traffic divided by the delivered bandwidth, and each of the four techniques improves one of the two. Striped packed streaming doubles the delivered bandwidth, from 21.9 to 43.9\,GB/s, at unchanged traffic, because a single master per projection half cannot fill a NoC port whereas balanced, prefetched streams keep all three ports busy. Quantization halves the traffic, from 5.14 to 2.64\,GB per token, at nearly unchanged bandwidth, and the latency halves with it. The two memory-side techniques account for $3.6\times$ of the $4.33\times$; quantization is also the only step with a significant arithmetic cost, the dequantization DSPs that raise the count from 1120 to 1589.
 
Rank reduction removes 13.9\% of the traffic and 12.9\% of the latency, so 93\% of the bytes it removes reach the board as time. Per layer type the match is similar: layers with a rank-896 input projection run 16\% faster than dense layers for 15.5\% fewer bytes, and layers with both projections factored 28\% faster for 31\% fewer, while dense layers are unchanged. This follows guidelines in Section~\ref{sec:projunit}: with the first factor shared by all four input-projection engines and the dense rows rebalanced, a low-rank layer takes about as long as its bytes take to arrive. The size of the 13.9\% itself is set by the operand audit of Section~\ref{sec:operands}, which keeps the gate and state rows and 56 of the 64 output projections dense; a larger gain would require a larger byte reduction or a higher delivered bandwidth rather than a change of the datapath.
 
The fused scan is the one compute-side step. It saves 4.6\% of latency and 122 UltraRAMs at unchanged traffic by shortening the on-chip chain between the two streaming phases, which raises the token-averaged delivered bandwidth from 39.9 to 42\,GB/s, 81\% of what the platform provisions.

\begin{figure}[t]
  \centering
  \includegraphics[width=\columnwidth]{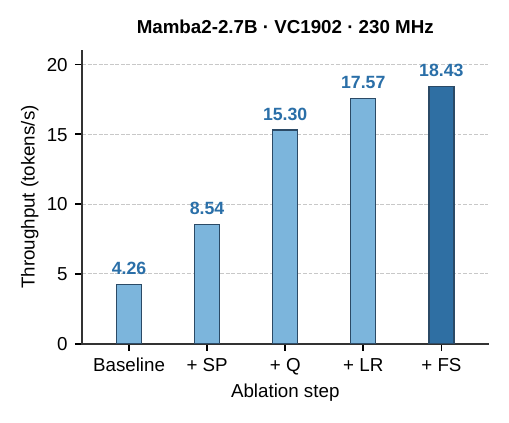}
  \caption{Decode throughput of each ablation step on the VC1902 from Table~\ref{tab:ablation}.}
  \label{fig:ablation-thpt}
\end{figure}

%% file: sections/06-Conclusion.tex
\section{Discussions and Limitations}

\textbf{Generalization beyond the Mamba family.} We applied uniform-rank truncated SVD at a 20\% reduction of the targeted operand to the Mamba projections of two hybrid models, Falcon-H1 0.5B~\cite{zuo2025falconh1familyhybridheadlanguage} and Granite 350M~\cite{granite} (Table~\ref{tab:cross-family}). Low-rank compression of the projections extends beyond Mamba2-2.7B, but the operand that tolerates it is model-dependent: on both hybrids the output projection compresses more gracefully than the $X$ slice of the input projection, the reverse of Mamba2-2.7B, and on Granite the $X$ slice is catastrophic. The operand audit of Section~\ref{sec:operands} therefore has to be repeated per model; the audit and the search are what transfer, not a particular allocation.

\textbf{Recovering quality at a fixed schedule.} Section~\ref{sec:searchresults} shows that whitened truncation recovers about a third of the perplexity penalty of plain SVD at the same ranks without any training. Light calibration or fine-tuning of the factors would recover more and allow lower ranks at the same ceiling; it is outside our post-training, fine-tuning-free setting and is a direction for future work.

\textbf{ASIC portability and the memory wall.} The absolute wall depends on memory technology. HBM or large SRAM raises the bandwidth ceiling and shifts the roofline crossover upward, but the relative benefit of structural rank reduction persists wherever projection
weight movement dominates per-token traffic, since fewer bytes must be moved regardless of how fast they move. On a memory-rich ASIC the benefit shrinks toward the crossover; on a bandwidth-constrained ASIC it is preserved or amplified. We also plan to implement rank reductions with other ASIC Mamba accelerators in the future.

 \begin{table}[t]
    \caption{Cross-family low-rank sensitivity (Wikitext-2 PPL, lower better; uniform 20\% reduction)}
    \centering
    \begin{tabular}{c|c|c|c}
    \toprule
        Model & Baseline & in\_proj\_x SVD & out\_proj SVD \\
        \midrule
         Falcon-H1 0.5B~\cite{zuo2025falconh1familyhybridheadlanguage} & 19.51 & 20.22 & 20.14 \\
         Granite 350M~\cite{granite} & 21.23 & 109.97 & 24.17 \\
         \bottomrule
    \end{tabular}

    \label{tab:cross-family}
    \vspace{-5mm}
\end{table}

\section{Conclusion}
This paper presented \textbf{LowRank-SSM}, a hardware-software co-design framework for rank-reduced Mamba-2 inference built around three coupled decisions: where to allocate rank, how to execute arbitrary rank schedules, and how to deliver the resulting weight streams efficiently. On the software side, sensitivity analysis identifies approximation-tolerant operands and a perplexity-constrained greedy search allocates rank per projection. On the hardware side, the projection datapath reuses the existing GEMV engines for both factor stages and exposes rank through loop bounds, while the platform-aware memory subsystem balances and prefetches the resulting streams. The resulting design sustains 42\,GB/s, with 93\% of the traffic removed by rank reduction appearing as latency reduction. On a Versal VC1902, the complete system reaches 16.79 tokens/s and 0.41\,J/token. The central result is that low-rank compression becomes an effective FPGA optimization only when rank allocation, rank-configurable execution, and memory delivery are designed together.